\documentclass[prb,twocolumn,superscriptaddress,showpacs,letterpaper]{revtex4}
\usepackage{graphicx}
\begin{document}
\newtheorem{theorem}{Theorem}
\newtheorem{acknowledgement}[theorem]{Acknowledgement}
\newtheorem{algorithm}[theorem]{Algorithm}
\newtheorem{axiom}[theorem]{Axiom}
\newtheorem{claim}[theorem]{Claim}
\newtheorem{conclusion}[theorem]{Conclusion}
\newtheorem{condition}[theorem]{Condition}
\newtheorem{conjecture}[theorem]{Conjecture}
\newtheorem{corollary}[theorem]{Corollary}
\newtheorem{criterion}[theorem]{Criterion}
\newtheorem{definition}[theorem]{Definition}
\newtheorem{example}[theorem]{Example}
\newtheorem{exercise}[theorem]{Exercise} 
\newtheorem{lemma}[theorem]{Lemma}
\newtheorem{notation}[theorem]{Notation}
\newtheorem{problem}[theorem]{Problem}
\newtheorem{proposition}[theorem]{Proposition}
\newtheorem{remark}[theorem]{Remark}
\newtheorem{solution}[theorem]{Solution}
\newtheorem{summary}[theorem]{Summary}   
\def\r{{\bf{r}}}
\def\j{{\bf{j}}}
\def\m{{\bf{m}}}
\def\k{{\bf{k}}}
\def\kt{{\tilde{\k}}}
\def\mt{{\hat{t}}}
\def\mG{{\hat{G}}}
\def\mg{{\hat{g}}}
\def\mGa{{\hat{\Gamma}}}
\def\mS{{\hat{\Sigma}}}
\def\mT{{\hat{T}}}
\def\K{{\bf{K}}}
\def\P{{\bf{P}}}
\def\q{{\bf{q}}}
\def\Q{{\bf{Q}}}
\def\p{{\bf{p}}}
\def\x{{\bf{x}}}
\def\X{{\bf{X}}}
\def\Y{{\bf{Y}}}
\def\F{{\bf{F}}}
\def\G{{\bf{G}}}
\def\bG{{\bar{G}}}
\def\mbG{{\hat{\bar{G}}}}
\def\M{{\bf{M}}}
\def\V{\cal V}
\def\tchi{\tilde{\chi}}
\def\tx{\tilde{\bf{x}}}
\def\tk{\tilde{\bf{k}}}
\def\tK{\tilde{\bf{K}}}
\def\tq{\tilde{\bf{q}}}
\def\tQ{\tilde{\bf{Q}}}
\def\si{\sigma}
\def\ep{\epsilon}
\def\hep{{\hat{\epsilon}}}
\def\al{\alpha}
\def\be{\beta}
\def\ep{\epsilon}
\def\bep{\bar{\epsilon}_\K}
\def\up{\uparrow}
\def\de{\delta}
\def\De{\Delta}
\def\up{\uparrow}
\def\dwn{\downarrow}
\def\ksi{\xi}
\def\etha{\eta}
\def\product{\prod}
\def\goto{\rightarrow}
\def\switch{\leftrightarrow}
\title{s-wave superconductivity phase diagram in the inhomogeneous two-dimensional attractive Hubbard model}

\author{K.~Aryanpour}
\affiliation{Department of Physics, University of California, 
Davis, California 95616} 
\affiliation{Department of Physics, SUNY at Buffalo, 
Buffalo, NY 14260} 

\author{T.~Paiva} 
\affiliation{Departamento de Fisica dos S\'olidos,
Instituto de F\'\i sica, Universidade Federal do Rio de Janeiro, 
Cx.P. 68528, 21945-970, Rio de Janeiro, RJ, Brazil}

\author{W.~E.~Pickett}
\affiliation{Department of Physics, University of California, 
Davis, California 95616}

\author{R.~T.~Scalettar} 
\affiliation{Department of Physics, University of California, 
Davis, California 95616}

\date{\today}
\begin{abstract}
We study s-wave  superconductivity in the two-dimensional square lattice
attractive Hubbard Hamiltonian for various inhomogeneous patterns of
interacting sites. Using the  Bogoliubov-de Gennes (BdG) mean field
approximation, we obtain the phase diagram for inhomogeneous patterns
in which the on-site attractive interaction $U_i$ between the electrons
takes on two values, $U_i=0$ and $-U/(1-f)$  (with $f$ the concentration
of non-interacting sites) as a function of average electron occupation
per site $n$, and study the evolution of the phase diagram as $f$
varies. In certain regions of the phase  diagram, inhomogeneity results
in a larger zero temperature average pairing amplitude (order parameter)
and also a higher superconducting (SC) critical temperature $T_c$,
relative to a uniform system with the same mean interaction strength
($U_i=-U$ on all sites).  These effects are observed for stripes,
checkerboard, and even random patterns of the attractive centers,
suggesting that the pattern of inhomogeneity is unimportant.  The phase
diagrams also include regions where superconductivity is obliterated due
to the formation  of various charge ordered phases. The enhancement of
$T_{c}$ due to inhomogeneity is robust as long as the electron doping
per site $n$ is less than twice the fraction of interacting sites
[$2(1-f)$] regardless of the pattern.  We also show that for certain
inhomogeneous patterns, when $n = 2(1-f)$,  increasing temperature can
work against the stability of  existing charge ordered phases for 
large $f$ and as a result, enhance $T_{c}$.   
\end{abstract}
%
\maketitle 
\section{Motivation}
\label{sec:introduction}            
\par Fascination with inhomogeneous superconducting (SC) phases extends
back several decades, with many conferences and monographs having been
devoted to the subject.\cite{inhomog.scy.conf,monograph2,dagotto94} For
conventional superconductors, the inhomogeneities were extrinsic, arising
from a granular nature of samples or due to the deliberate synthesis of
disordered built materials or films. The high temperature
superconductors (HTS) introduced new aspects into this area of study.
First, inhomogeneous states (normal and SC) seem to be intrinsic to HTS,
at least in the underdoped regime, similar to quenched disorder in the
metal-insulator transition in two dimensions.\cite{Lee85,Belitz94}
Secondly, the inhomogeneity occurs on a smaller length scale of just a
few lattice constants as evidenced by scanning tunneling spectroscopy
(STS) at the nanoscale.\cite{cren,pan,howald,lang} 

\par In addition, the strong electronic correlation in HTS cuprates
plays a major role in the elucidation of the inhomogeneous SC state;
indeed the inhomogeneity  is widely believed to arise from the strong
intra-atomic interactions that tend to frustrate bandlike  conduction,
to induce local magnetic moments, and to drive charge and spin order on
a few lattice constant  scale.  The study of the doped Mott insulating
phase has been one of the most active areas of theoretical  study of
HTS, one that has proven to be unexpectedly complex and rich and which
is still under intense  exploration and debate. Surprisingly, holes
doped into the high temperature superconductors tend to arrange
themselves non-uniformly in the CuO$_{2}$ planes in the form of stripes,
checkerboard or perhaps even more exotic
structures.\cite{mcelroy,hanaguri,vershinin,mook,tranquada} Moreover,
spatially varying density and spin structures have also been observed in
the physics of the manganites \cite{renner,burgy,dagotto99,dagotto01} and
cobaltites.\cite{foo,lee}

\par A variety of physically relevant models such as the repulsive
Hubbard and t-J Hamiltonians have been extensively studied to understand
the interplay between spatial inhomogeneity, magnetism, and
superconductivity.\cite{zaanen,machida,kato,white,vojta,seibold,kivelson,huai,maska,tsai,seibold2,garg,zhang,robertson,ghosal,dobrosavljevic92,dobrosavljevic94,dobrosavljevic03,yukalov1,yukalov2}
In the repulsive Hubbard and t-J Hamiltonians in particular,
inhomogeneity has been introduced either through the hopping amplitude
$t$ or magnetic coupling $J$ or the local energy on the lattice sites.
For the $2D$ square lattice these two models are known to display
antiferromagnetism at half-filling, and, although it is less certain,
perhaps also d-wave superconductivity when doped.  There is considerable
evidence that they also might possess inhomogeneous stripe or
checkerboard ground
states.\cite{white,vojta,seibold,kivelson,huai,tsai,seibold2,zhang,robertson}.
Phenomenological d-wave BCS Hamiltonians with spatially inhomogeneous
pairing amplitude \cite{valdez,seo,andersen} or lattice site energy
\cite{caixeiro,andersen} have also been employed mostly to reproduce the
local density of states (LDOS) results obtained from scanning tunneling
microscopy (STM).\cite{jamei} In addition, there have been theoretical 
studies of the SC quantum phase fluctuations using the QED$_{3}$ effective theory of the HTS in relation with the inhomogeneous pattern formation in cuprates from the STM experimental results.\cite{melikyan,tesanovic}      

\par While DMRG treatments \cite{white} provide detailed information on
the real space charge, spin, and pairing orders, the precise nature of
the interplay, and whether the different orders compete or cooperate,
remains unclear. In addition, the enhancement of the superconducting
transition temperature $T_c$ by local inhomogeneity has been
demonstrated by Martin {\it et al.} in Ref. [\onlinecite{martin}] and
also in 
Ref. [\onlinecite{olderwork}].
Recently, the $XY$ model Hamiltonian with certain types of inhomogeneous
patterns for the coupling constant between spins sitting on two nearest
neighboring sites has also been shown to enhance $T_{c}$ by Loh {\it et
al.} in Ref. [\onlinecite{loh}].  

\par Many of the basic characteristics of this short-range-scale
inhomogeneous superconducting state can be addressed with a more
tractable model, one which is well understood in the homogeneous limit.
This model is the {\it attractive} Hubbard model, which has been applied
previously to address some aspects of the impact of inhomogeneity.
Recently old suggestions based on ``negative U'' superconductivity have
been revived,\cite{olderwork,chatterjee} which may provide additional
applications for the results we present in this paper. Tl-doped PbTe
achieves a critical temperature up to 1.5 K, and more extensive heat
capacity and transport data\cite{stanford1,stanford2} have led to an
analysis in terms of a ``charge Kondo effect'' that could be linked to
the observed superconductivity.\cite{dzero}  This system intrinsically
involves both negative U centers and inhomogeneity. 

\par  This article extends previous work \cite{olderwork} to a more
general range of non-interacting site concentration $f$ values.  We show
the presence of different conduction phases both in the phase diagram at
$T=0$ and in the density of states (DOS). We also show the local
occupation and SC order parameter for electrons on different lattice
sites as the concentration $f$ varies for different inhomogeneity
patterns. Finally, the $T_{c}$ enhancement conditions are also extracted
based on the relationship between the average doping of electrons $n$ on
the lattice and inhomogeneity concentration $f$.

\par The paper is organized as follows: in the next section we introduce
our model and describe the method we have employed. In Sec. III we
present and discuss the phase diagram at zero temperature. 
Sec. IV contains our finite temperature results, and Sec. V summarizes our
findings.

\section{Model and Methodology}
\label{sec:formalism}

\par This article focuses on a general question: ``Under what conditions
is it more favorable to have an inhomogeneous pairing attraction,
compared to the same average strength
spread homogeneously over the lattice?'' By `conditions'  we   refer to
the average occupation number of electrons per site $n$, the average
attraction strength  per lattice site $\bar U$, which remains constant
in any comparison between systems with and without inhomogeneity, and
the inhomogeneity concentration $f$.  We address this question by
comparing the  average zero temperature pairing amplitude $\bar\Delta$
over the entire lattice and the SC transition  temperature $T_{c}$ for a
system in the presence and absence of inhomogeneity. 

For the cuprate superconductors, for example, such a question is
complicated  by the presence of other types of order such as
antiferromagnetism, exotic spin-gap phases, and by the nontrivial d-wave
symmetry of the SC order parameter. For these systems and phenomena,
models like the repulsive  Hubbard or t-J Hamiltonians are
essential.\cite{zaanen,machida,kato,white,vojta,seibold,kivelson,huai,maska,tsai,seibold2,garg,zhang,robertson,ghosal,dobrosavljevic92,dobrosavljevic94,dobrosavljevic03}
Nevertheless, it is yet beneficial to study the problem first by
employing a more simple and phenomenological model. Here we will present
a solution of the inhomogeneous Bogoliubov-de Gennes (BdG) equations for
the attractive Hubbard Hamiltonian,
\begin{eqnarray} \label{eq:attr-hub-mod}
H=&-&t\sum_{<ij>,\sigma}(c^{\dag}_{i\sigma}c_{j\sigma} +
c^{\dag}_{j\sigma}c_{i\sigma}) \nonumber \\ &-&
\mu\sum_{i\sigma}c^{\dag}_{i\sigma}c_{i\sigma}-\sum_{i}\big|U_{i}\big|
n_{i\uparrow}n_{i\downarrow}\,, \end{eqnarray}
with $t$ the hopping amplitude, $\mu$ the chemical potential and $U_{i}$
the local attractive interaction between the fermions of opposite spins
residing on the same lattice site $i$. We will study inhomogeneous
patterns in the interaction $U_{i}$. The origin of the attraction in
this model can result from, for example, integrating out a local phonon
mode.\cite{micnas90} The two-dimensional uniform attractive Hubbard
model is known to yield degenerate superconductivity and charge density
wave (CDW) long range order at half-filling and zero
temperature.\cite{robaszkiewicz,shiba,emery} However, away from
half-filling, the CDW pairing symmetry is broken and superconductivity
is more favorable, and the SC phase transition is at finite temperature.

\par The BdG mean field decomposition bilinearizes the Hamiltonian by
replacing the local pairing amplitude and local density by their average
values, $\Delta_{i}=\big<c_{i\uparrow}c_{i\downarrow}\big>$ and
$\big<n_{i\sigma}\big>=\big<c^{\dag}_{i\sigma}c_{i\sigma}\big>$ and
yields the quadratic effective Hamiltonian
\begin{eqnarray} \label{eq:Heff-bdg}
{\cal{H}}_{eff}=&-&t\sum_{<ij>,\sigma}(c^{\dag}_{i\sigma}c_{j\sigma} +
c^{\dag}_{j\sigma}c_{i,\sigma}) - \sum_{i\sigma}{\tilde\mu_i}
c^{\dag}_{i\sigma}c_{i\sigma} \nonumber \\ &-& \sum_{i}
\big|U_{i}\big|\big[\Delta_{i}c^{\dag}_{i\uparrow}c^{\dag}_
{i\downarrow}+\Delta^{*}_{i}c_{i\downarrow}c_{i\uparrow}\big]\,,
\end{eqnarray}
where ${\tilde\mu_i}=\mu+\big|U_{i}\big| \langle n_{i} \rangle/2$
includes a site-dependent Hartree shift with $\langle n_{i} \rangle
=\sum_{\sigma} \langle n_{i\sigma} \rangle$.  All energies will be
referenced to $t=1$. 
\begin{figure} \includegraphics[width=3.2in]{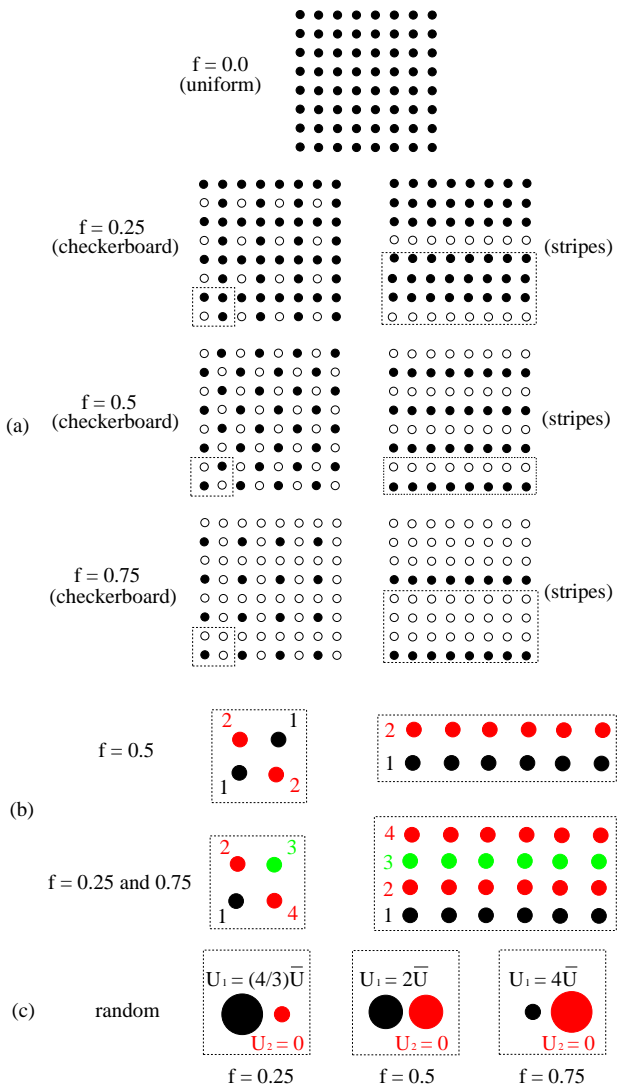}
\caption[a]{(Color online) Panel (a): Regular patterns for the
interacting sites in the attractive Hubbard model at different
inhomogeneity concentration values. Stripes and checkerboard have
been particularly selected because of their relevance to the
experimental observations in cuprates. Panel (b): Color coding and
numbering different types of sites for the checkerboard and stripes
blocks as presented in panel (a) (two colors for $f=0.5$ and four for
$f=0.25$ and $0.75$). Sites carrying identical color code and number are
equivalent by the symmetry in the lattice geometry. Panel (c): Color
coding of the lattice sites for the random inhomogeneous pattern at
different $f$ values. Regions of interacting (non-interacting) sites are
coded black (red or dark gray in the grayscale version) with the appropriate weight of $1-f$ ($f$).}
\label{patterns} \end{figure}

\par We adopt the criterion of comparing  the tendency for
superconductivity in the homogeneous system with the same attraction
$-U$ on all lattice  sites, with cases when sites with attraction are
mixed with sites where the attraction is absent, i.e.,  $U_{i}=0$.
\cite{martin,litak,hurt} Specifically, we have studied systems in which
sites with attractive  interaction are randomly distributed \cite{litak}
or arranged in checkerboard and stripe patterns.  The last two regular
patterns have been purposely chosen due to their relevance to the
experimentally observed pattern formation in the HTS cuprates. 

Fig. \ref{patterns}, panel (a),  presents the patterns for the
interacting lattice sites with four different values for the fraction of
non-interacting sites $f$. The uniform pattern corresponds to $f=0$ with
interaction $U_{i}=\bar  U=-U$ on all lattice sites. Checkerboard,
stripes and random patterns with $f=0.25$ include $1-f=75\%$ interacting
sites  with $U_{i}=\frac{4}{3}\bar U=-\frac{4}{3}U$ and $f=25\%$
non-interacting sites with $U_{i}=0$.
For $f=0.5$, half
of the sites are interacting with $U_{i}=-2U$ and half non-interacting
with $U_{i}=0$. 
$f=0.75$ corresponds  to $25\%$ interacting sites with
$U_{i}=-4U$ and $75\%$ non-interacting sites with $U_{i}=0$  once again
averaging to $\bar U=-U$ per site. 

For the random pattern we have averaged over typically 20 different
disorder realizations.
One may note that regardless of the pattern and the impurity fraction,
the average attraction per site,   i.e., $\bar U=-U$ remains constant.
We adopt this criterion throughout the rest of this article for
comparison between uniform and inhomogeneous lattices. This criterion is
believed to be most appropriate  for exhibiting the effect of
inhomogeneity in particular in the systems having the same strength on
average for forming superconducting Cooper pairs. 

Panel (b) depicts the color coding and numbering  of the lattice sites
for the checkerboard and stripes based on their value of $f$. Due to the
regular  geometry and periodicity of their inhomogeneous patterns,
lattices with the checkerboard and stripe  patterns can be classified
into blocks including two and four different types of sites at $f=0.5$
and  $f=0.25$ or $0.75$ respectively as illustrated by different colors
and numbers in panel (b) of Fig.  \ref{patterns}. Sites carrying the
same color or number are equivalent by the symmetry in the pattern
geometry. For the random pattern, due to the lack of both regularity and
periodicity, 
we average over all the interacting and all the non-interacting sites
separately (black and red (Red shows dark gray in the grayscale version) 
regions in panel (c) of Fig.\ref{patterns} respectively) before the configurational averaging over all  different impurity patterns is performed. 
%

\par We self-consistently diagonalize the  BdG mean-field Hamiltonian in Eq.
\ref{eq:Heff-bdg} by assigning initial values  to the
local occupation number $n_{i}$ and order parameter $\Delta_{i}$ and
solving again for these parameters after
diagonalizing the Hamiltonian, until convergence is achieved  at a
desired tolerance. For the checkerboard and striped patterns, sites with
the same color have the same density, due to symmetry, and do not need
to be averaged.  For the random pattern we calculate the average
occupation number $ n_{\rm color}$ and order parameter $\Delta_{\rm
color}$ per site at the different types of sites by averaging
\begin{eqnarray} \label{eq:avg-occp-delta-reg} n_{\rm color} =
\frac{1}{N_{\rm color}}\sum_{i\in\{\rm color\}}n_{i}\nonumber \\
\Delta_{\rm color} = \frac{1}{N_{\rm color}}\sum_{i\in\{\rm
color\}}\Delta_{i} \end{eqnarray}

For the checkerboard and striped patterns with $f=0.5$ and for the
random pattern (all $f$) we simply have $n_{\rm attrac}=  n_{\rm black}$
and $ n_{\rm free}=n_{\rm red}$ (Red shows dark gray in the grayscale version). For the striped and checkerboard with $f=0.25$ we have $n_{\rm attrac}=  ( n_{\rm green}+ 2   n_{\rm red})/3$ (Green shows light gray in the grayscale version) and $ n_{\rm free}= n_{\rm black}$, whereas for $f=0.75$ we have $n_{\rm attrac}= n_{\rm black}$ and $ n_{\rm free}=( n_{\rm green}+ 2n_{\rm red})/3 $. 
The same combinations hold for ${\Delta}_{\rm attrac}$ and ${\Delta}_{\rm free}$.  The average occupation number $n$ and order parameter $\bar \Delta_{\rm inhom}$ per site are defined 
\begin{eqnarray} \label{eq:avg-occp-delta-ireg} n = (1-f){ n}_{\rm
attrac}+f{ n}_{\rm free}\nonumber \\ \bar\Delta_{\rm inhom} =
(1-f){\Delta}_{\rm attrac}+f{\Delta}_{\rm free} \end{eqnarray}
The chemical potential $\mu$ in Eq. \ref{eq:Heff-bdg} is
self-consistently adjusted after every iteration in order to arrive at a
desired total average occupation per site $n$ for the entire lattice.
For the regular patterns, i.e., uniform, checkerboard and stripes, due
to their periodicity, by Fourier transforming the Hamiltonian into
momentum space, we significantly reduce the numerical cost of the
calculations and at the same time can increase the size of the lattice
close to the thermodynamic limit to avoid finite size artifacts in the
results (up to $1500\times1500$ lattice sites in our calculations). For
the random pattern, however, such a simplification is not possible due
to the lack of periodicity. Hence, we are limited to the finite size
lattices of up to $24\times24$ sites. As a result, especially at small
values for the average on-site interaction magnitude $|\bar U|$, finite
size effects are to be cautiously monitored. Our calculations also
include the density of states (DOS) for the entire lattice. We study
simultaneous variations of the average on-site interaction magnitude
$|\bar U|$, occupation number $n$ and also temperature and their effects
in the average order parameter $\bar \Delta_{inhom}$ and DOS. Our goal
is to obtain the phase diagram for the effect of inhomogeneity in
superconductivity and discuss the conditions under which inhomogeneity
can result in enhancements in the average  superconductivity order
parameter or SC phase transition temperature $T_{c}$.   

\par It is further realized that our conventional mean-field  approach
does not capture the Kosterlitz-Thouless nature of the phase transition
in two  dimensions. Nevertheless, this weakness can be repaired
\cite{mayr} upon regarding the local pairing  amplitudes as complex
variables and performing a finite temperature Monte Carlo integration
over the  associated amplitude and phase degrees of freedom. Unlike BCS,
this Monte Carlo mean field (MCMF)  approach allows identification of
the weak and strong coupling regimes via the phase correlation
function. In an earlier work \cite{olderwork} this Monte Carlo technique
was employed as an independent  examination for the validity of our
results and the agreement between the two techniques was clearly
confirmed. 
%
%
\section{Phase Diagram at $T=0$} \label{sec:resultsT0}

\par
\begin{figure*} \includegraphics[width=6.8in]{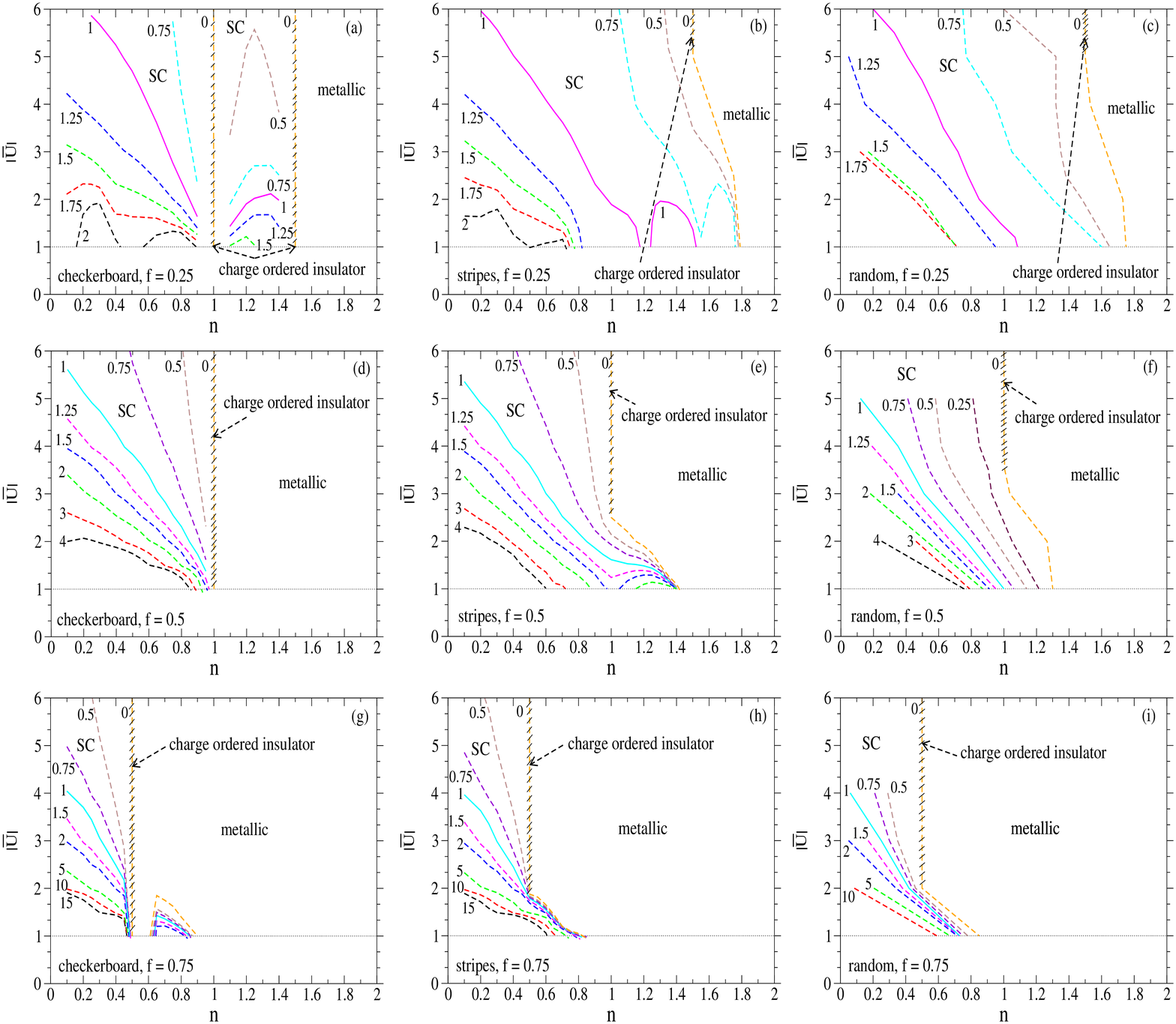}
\caption[a]{(Color online) Panel (a): The contour plot phase diagram for
the checkerboard pattern at $f=0.25$ and $T=0$. The horizontal axis
presents the average occupation of electrons per site $n$ and the
vertical axis refers to the average interaction magnitude between two
electrons per site $|\bar U|$.  Lines with numbers next to them
correspond to different ratios of $r=\bar\Delta_{\rm
checkerboard}/\Delta_{\rm uniform}$. Solid lines at $r=1$ determine the
enhancement boundary. Dashes along $r=0$ lines indicate charge ordered
insulating phase behavior. Dotted lines at $|\bar U|=t$ are lower limits
for the interaction as for too small $|\bar U|$ values, $r$ is an ill
defined quantity. Panel (b): The same results for stripes with
$r=\bar\Delta_{\rm stripes}/\Delta_{\rm uniform}$ at $f=0.25$ and $T=0$.
$r=0$ line for stripes is diverted towards larger $n$ values at smaller
$|\bar U|$ and does not run down to arbitrarily small $|\bar U|$ values
at $n=1.5$. Dashes along $r=0$ line for stripes appear only at $n=1.5$
and beyond that, $r=0$ corresponds to a metallic phase. Panel (c)
presents results for the random pattern. Similar to the stripes, for the
random pattern, $r=0$ line does not run down to arbitrarily small $|\bar
U|$ values at $n=1.5$ either. Panels (d), (e) and (f) correspond to
$f=0.5$. Note that the charge ordered phases    for the striped and
random patterns at $f=0.5$ again occur only at $n=1$ portion of the
$r=0$ line and beyond that, $r=0$ yields a metal. Similarly in panels
(g), (h) and (i) corresponding to $f=0.75$, all three different
inhomogeneous patterns have a $r=0$ line portion above $n=0.5$ at which
the systems become metallic. Panel (i) also lacks the $r=15$ contour due
to the finite size effect uncertainties at small $|\bar U|$ values.}
\label{phasediag} \end{figure*}

 
Fig.~\ref{phasediag} presents  the phase diagram for the average
interaction magnitude $|\bar U|$ and electron doping $n$ per site  at
$T=0$ for three different inhomogeneous patterns  of checkerboard,
stripes, and random, and for $f$ values of $0.25$, $0.5$ and $0.75$.  
We show isocontours of $r=\bar\Delta_{\rm
inhom}/\Delta_{\rm uniform}$, i.e., the ratio of the average
inhomogeneous pattern order parameter over its uniform pattern
counterpart.  Thus, whenever $r>1$, inhomogeneity leads to a larger
average order parameter at $T=0$ compared to a  homogeneous system  and
therefore is more favorable for superconductivity over a uniform pattern
of the  interacting sites. We also adopt the lower limit of $|\bar U|=t$
since for smaller values of $|\bar U|$, $r$ will be the ratio of two
very small numbers and is subject to numerical uncertainty. The first
row of Fig.  \ref{phasediag} (panels (a)-(c)) corresponds to the
concentration value of $f=0.25$ for the non-interacting sites. At first
glance, one can observe that regardless of the geometry for the
inhomogeneity, above $n=1.5=2(1-f)$, inhomogeneity gradually (or
abruptly for the checkerboard pattern in panel (a)) results in the
obliteration of superconductivity consistent with the findings of Litak
{\it et al}\cite{litak}. We can understand how this obliteration takes
place if we examine the behavior of the system in strong coupling. When
we start with an empty system and  add electrons they are placed on  the
attractive sites first due to the strong attractive interactions. It is
useful to define $n^*=2(1-f)$, which for a given $f$ corresponds to the
density for which all attractive sites are doubly occupied  and all free
ones are empty.  As we will see below, this density corresponds to an
insulating charge ordered state.  Above this density, superconductivity
is reduced because the pairs cannot move within the attractive
sublattice, since it is completely filled. 

\par For the checkerboard pattern in panel (a), there are two insulating
regions within the phase diagram at $n=1$ and $1.5$ (hatched orange
lines) both corresponding to the formation of charge ordered phases of
electrons in the interacting sites. No superconductivity was observed
for $n=1$ and $1.5$ down to the lower limit of $|\bar U|=t$. Beyond
$n=1.5$, the system becomes metallic. For stripes as shown in panel (b),
similar features as in panel (a) are observed. One exception is the lack
of the charge ordered insulating phase at $n=1$. This can be the result
of further overlap between the Cooper pairs since for the stripes,
nearest neighboring sites are interacting in one dimension. The charge
ordered insulating phase at $n=1.5$ also forms for rather higher $|\bar
U|$ values compared to the checkerboard pattern. The random pattern in
panel (c) also shows features similar to the stripes. 

The second row (panels (d)-(f)) corresponds to $f=0.5$ with rather
similar features to the first row. As anticipated, superconductivity
gradually goes away above $n=1=n^*$ for all the patterns. For the
checkerboard in panel (d), and $n>1$, superconductivity strictly goes
away and the system turns metallic. For the striped and random patterns
however, superconductivity persists slightly above $n=1$ until it is
totally obliterated. At $n=1$, all three inhomogeneous patterns exhibit
a charge ordered insulating phase  for large enough values of $|\bar U|$
(or the smallest value of $|\bar U|$ for the checkerboard).
Nevertheless, it can be readily seen that for $f=0.5$ compared to
$f=0.25$, the enhancement of the average order parameter due to
inhomogeneity is considerably larger as the ratio $r$ increases for
small $|\bar U|$ values. 

The third row (panels (g)-(i)) for $f=0.75$ shows ratios as large as
$r=15$ for small values of $|\bar U|$ and $n$ values not much larger
than $0.5=n^*$. For $f=0.75$ also, superconductivity gradually dies away
when $n>0.5=n^*$ and a charge ordered phase sets in for large enough
$|\bar U|$ values at $n=0.5=n^*$ analogous to $f=0.25$ and $0.5$. The
only difference is a slight remnance of superconductivity for the
checkerboard pattern at $n>0.5$. Thus, by further diluting the
interacting sites in the lattice and keeping the attractive pairing
energy constant at the same time, superconductivity is driven towards
smaller $n$ values; on the other hand, the average order parameter
becomes significantly more enhanced due to inhomogeneity. Generally,
regardless of the pattern, for large enough $|\bar U|$ values,
inhomogeneity weakens superconductivity for every value of $n$ due to
the localization and compression of the Cooper pairs in the interacting
sites. For $n<n^*$, ${\bar\Delta}_{\rm inhom}$ increases as a function
of $|\bar U|$ and saturates for large $|\bar U|$ values.  For $n\geq
n^*$,  ${\bar\Delta}_{\rm inhom}$ reaches a maximum as a function of
$|\bar U|$ and eventually vanishes for large enough $|\bar U|$ values.
However, ${\Delta}_{\rm uniform}$ is a monotonically increasing function
of $|\bar U|$ and is symmetric with respect to $n=1$. Therefore, for
sufficiently large $|\bar U|$, the ratio $r=\bar\Delta_{\rm
inhom}/\Delta_{\rm uniform}$ becomes less than one while $n<n^*$ and
eventually zero when $n\geq n^*$ as illustrated in Fig. \ref{phasediag}.    
\begin{figure*} \includegraphics[width=6.8in]{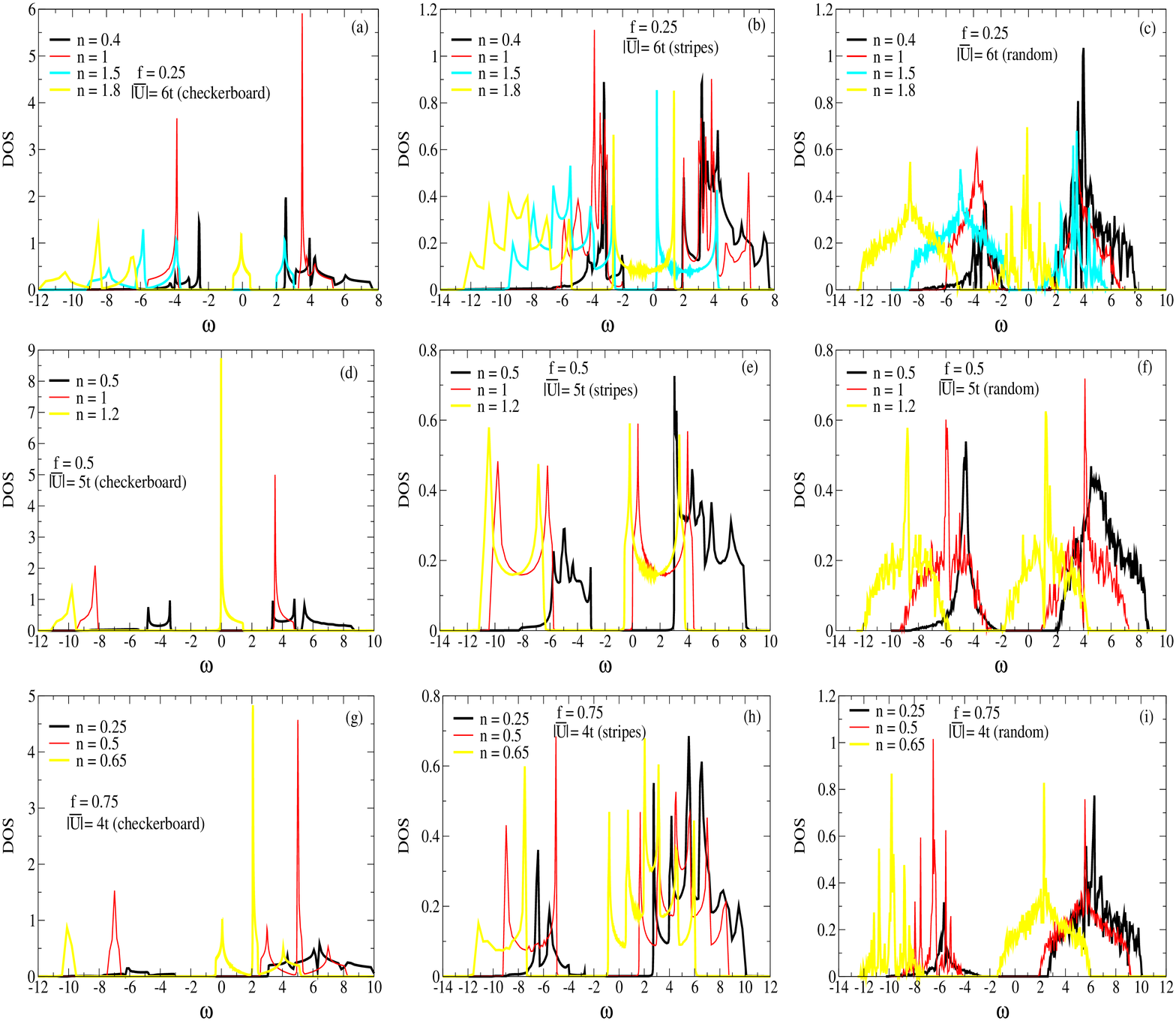}
\caption[a]{(Color online) Panel (a): Density of state (DOS) for the
checkerboard pattern at $f=0.25$ and ${|\bar U|}=6t$ (the largest in our
calculations for $f=0.25$) for different values of the average electron
occupation per site $n$. Panel (b) and panel (c): The same results as in
panel (a) for striped and random patterns respectively. Panel (d):
Results of panel (a) at $f=0.5$ and ${|\bar U|}=5t$ (the largest in our
calculations for $f=0.5$). Panel (e) and (f): The same results as in
panels (b) and (c) for striped and random patterns respectively. Panels
(g), (h) and (i) correspond to $f=0.75$ and ${|\bar U|}=4t$ (the largest
in our calculations for $f=0.75$) for the checkerboard, striped and
random patterns respectively. The particular selection of colors is for better visibility in both the color and grayscale versions.} \label{dos.T0.0} \end{figure*}
\begin{figure*} \includegraphics[width=6.8in]{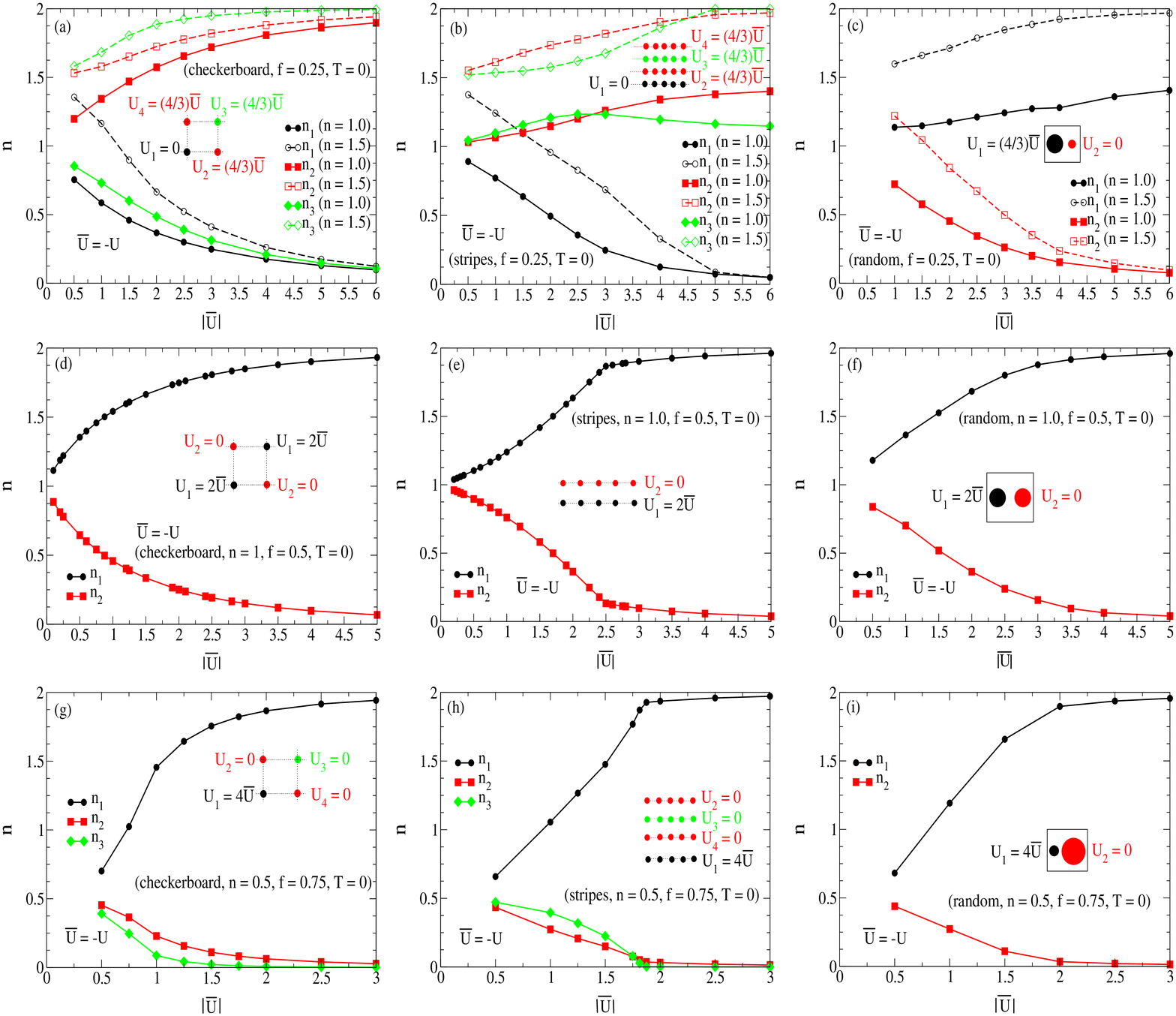}
\caption[a]{(Color online) Panel (a): The evolution of the local
electron occupation number $n_i$ on different lattice sites (As color
coded inside the blocks in Fig.\ref{patterns}, panels (b) and (c)) for
the checkerboard pattern at $f=0.25$ and $n=1$ and $n=1.5$ (referring to
charge ordered phases in Fig.\ref{phasediag}, panel (a)) as a function
of $|\bar U|$. Panels (b) and (c): The same results as in panel (a) for
the striped and random patterns respectively. For the random pattern,
data are taken by averaging the occupation number over the interacting
and non interacting sites. Panels (d), (e) and (f): The same results at
$f=0.5$ and $n=1$ (charge ordered phases in Fig.\ref{phasediag}, panels
(d), (e) and (f)). Also panels (g), (h) and (i) at $f=0.75$ and $n=0.5$
(charge ordered phases in Fig.\ref{phasediag}, panels (g), (h) and
(i)).} \label{occup-insulate.T0.0} \end{figure*}
\begin{figure*} \includegraphics[width=6.8in]{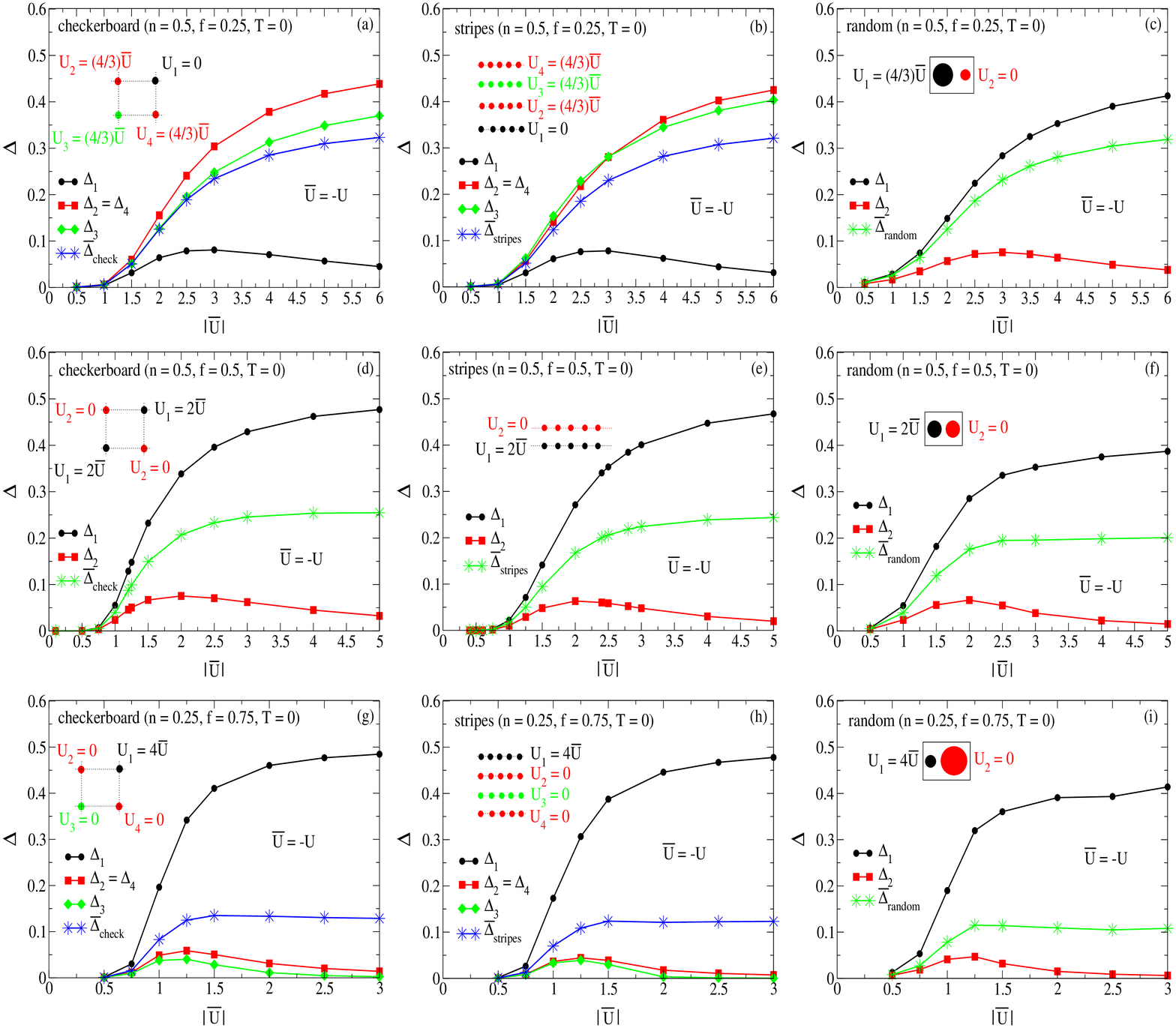}
\caption[a]{(Color online) Panel (a): The evolution of the local order
parameter $\Delta_i$ on different lattice sites (As color coded inside
the blocks in Fig.\ref{patterns}, panels (b) and (c)) for the
checkerboard pattern at $f=0.25$ and $n=0.5$ as a function of $|\bar
U|$. The curve symbolized with stars refers to the average order
parameter $\bar\Delta_{\rm inhom}$. Panels (b) and (c): The same results
as in panel (a) for the striped and random patterns respectively. One
notices that curves for $\Delta_2$ and $\Delta_3$ in panel (b) intersect
around ${|\bar U|}\approx 3t$. Panels (d), (e) and (f) present the same
results at $f=0.5$ and $n=0.5$ for the checkerboard, striped and random
patterns respectively. Also panels (g), (h) and (i) at $f=0.75$ and
$n=0.25$.} \label{delta.site.evolve} \end{figure*}

\par The obliteration of superconductivity is associated with the
vanishing of average order parameter $\bar\Delta_{\rm inhom}=0$. Whether
a non-SC state is a metal or insulator is determined by the DOS results
for that state. In Fig. \ref{dos.T0.0}, panel (a) presents the DOS
results for the checkerboard pattern at $|\bar U|=6t$ (the largest in
our calculations for $f=0.25$) and four different values for the average
electron doping $n$. There is a gap in the DOS around the Fermi energy
at $\omega=0$ at $n=0.4$. This gap corresponds to a SC state as for
$n=0.4$, ${\bar\Delta}_{inhom}\ne0$. The gaps at $n=1$ and $n=1.5$ both
correspond to insulating states as for both these $n$ values
${\bar\Delta}_{inhom}=0$. At $n=1.8$, we also find
${\bar\Delta}_{inhom}=0$ according to Fig.  \ref{phasediag}, panel (a).
However, the DOS at $n=1.8$ has a finite value around the Fermi energy
at $\omega=0$ as shown in panel (a) of Fig. \ref{dos.T0.0}, indicating a
metallic state. 

In panels (b) and (c) in Fig. \ref{dos.T0.0} for striped and random
patterns respectively, gaps at $n=1$ do not correspond to insulating
phases as opposed to panel (a) whereas they do correspond to insulators
for $n=1.5$. In the second row of Fig. \ref{dos.T0.0} (panels (d)-(f)),
for all the patterns at $f=0.5$ and $|\bar U|=5t$ (the largest in our
calculations for $f=0.5$), the system is superconducting for $n<1$,
insulating for $n=1$ and metallic for $n>1$ in conjunction with the
results of the second row in Fig. \ref{phasediag}. The third row of Fig.
\ref{dos.T0.0} (panels (g)-(i)), for $f=0.75$ and $|\bar U|=4t$ (the
largest of our calculations for $f=0.75$), confirms the results
presented in Fig. \ref{phasediag}, (panels (g)-(i) respectively),
namely, superconductivity for  $n<n^*$, insulator at $n=0.5=n^*$  and
large enough $|\bar U|$ and metal for $n=0.65>n^*$ for all inhomogeneous
patterns.

 \par The insulating state for the phase diagram in Fig. \ref{patterns}
at all values of $f$, $n$ and all inhomogeneous patterns is always
associated with the formation of a charge ordered state. In Fig.
\ref{occup-insulate.T0.0}, panel (a) for the checkerboard at $f=0.25$,
for the insulating phase at $n=1$, as $|\bar U|$ increases, electrons
form pairs in the interacting sites with higher symmetry [sites $2$ and
$4$ in red (dark gray in the graycsale version)], leaving the non-interacting and lower symmetry interacting
sites [sites $1$ and $3$ in black and green (light gray in the grayscale version), respectively] essentially
empty. For the insulating phase in panel (a) at $n=1.5$, the lower
symmetry interacting site [site $3$ in green (light gray in the grayscale version)] also obtains a pair
leaving only the non-interacting site (site $1$ in black) empty. In
other words, charges rearrange themselves into ordered pair
configurations forming an insulating phase. According to panel (b) in
Fig.  \ref{patterns} for stripes, $n=1$ does not correspond to an
insulating phase. Panel (b) in Fig.  \ref{occup-insulate.T0.0} confirms
this finding as the local occupation of the interacting sites (sites
$2$, $3$ and $4$ in red (dark gray in the grayscale version), green (light gray in the grayscale version) and red respectively) for large $|\bar
U|$ at $n=1$ does not approach $2$ while for the non-interacting sites
(sites $1$ in black) it approaches zero. 

For the random pattern in panel (c) of Fig. \ref{occup-insulate.T0.0} at
$f=0.25$, we plot $\bar n_{\rm black}$ and $\bar n_{\rm red}$ (Red shows dark gray in the grayscale version) as defined
in Eq. \ref{eq:avg-occp-delta-reg}. The fact that $\bar n_{\rm black}$
does not approach $2$ as $|\bar U|$ increases (no pair compression) is
consistent with the lack of an insulating phase at $n=1$ for the random
pattern at $f=0.25$ (Fig. \ref{patterns}, panel (c)). However, at
$n=1.5$, both striped and random patterns turn insulating as a result of
charge ordered phase formation as shown in panels (b) and (c) in  Fig.
\ref{occup-insulate.T0.0}, where their occupation numbers on the
interacting sites all approach $2$ as $|\bar U|$ increases. Charge
ordered phase formation is more obvious in the second row (panels
(d)-(f)) and third row (panels (g)-(i)) of Fig.
\ref{occup-insulate.T0.0} for $f=0.5$ and $f=0.75$ respectively at large
enough $|\bar U|$ values. 

\par The proximity effect for the non-interacting sites neighbored by
the interacting sites plays a key role in the magnitude of the average
order parameter ${\bar\Delta}_{\rm inhom}$ in the inhomogeneous lattice.
In Fig. \ref{delta.site.evolve}, panel (a), for the checkerboard at
$f=0.25$ and $n=0.25$, the local order parameter on all the interacting
sites ($2$, $3$, $4$ in red (dark gray in the grayscale version), green (light gray in the grayscale version) and red respectively) increases as a
function of $|\bar U|$. The non-interacting site ($1$ in black) is also
superconducting due to the proximity effect of its neighboring sites.
However, its local order parameter has a maximum at a critical $|{\bar
U}_{c}|$ value beyond which it decreases as a result of the compression
of the Cooper pairs in the interacting sites and therefore their weaker
overlap around the non-interacting site. Thus, ${\bar\Delta}_{\rm
inhom}$ on all these four sites will be larger than its uniform pattern
counterpart due to this proximity effect as long as the non-interacting
site local order parameter ($\Delta_1$ in this case) does not plummet
too far down with respect to its maximum as a function of $|\bar U|$. 

Panel (b) shows the same behavior for stripes. In panel (b), there is an
intersection between $\Delta_2$ and $\Delta_3$ near ${|\bar U|}\approx
3t$ indicating that due to the particular symmetry  of  the stripes,
sites $2$ and $3$ behave very closely.  In panel (c), we have plotted  $
\Delta_{\rm black}$ and $ \Delta_{\rm red}$ (Red shows dark gray in the grayscale version) as defined in Eq.
\ref{eq:avg-occp-delta-reg} and it is clear that $ \Delta_{\rm red}$
eventually falls off at large $|\bar U|$ values. In the second row of
Fig. \ref{delta.site.evolve} (panels (d)-(f)) corresponding to $f=0.5$
and $n=0.5$, there are only two different sites for each pattern and the
lattice has a more dilute interacting pattern. As a result, compared to
$f=0.25$ results, ${\bar\Delta}_{\rm inhom}$ at $f=0.5$ tends to
saturate for large $|\bar U|$ values for all the patterns. In the third
row of  Fig.  \ref{delta.site.evolve} (panels (g)-(i)) for $f=0.75$, the
lattice is even more dilute in terms of interacting energy. Therefore,
${\bar\Delta}_{\rm inhom}$ shows even faster saturation at smaller
$|\bar U|$ values.    %
\begin{figure} \includegraphics[width=2.4in]{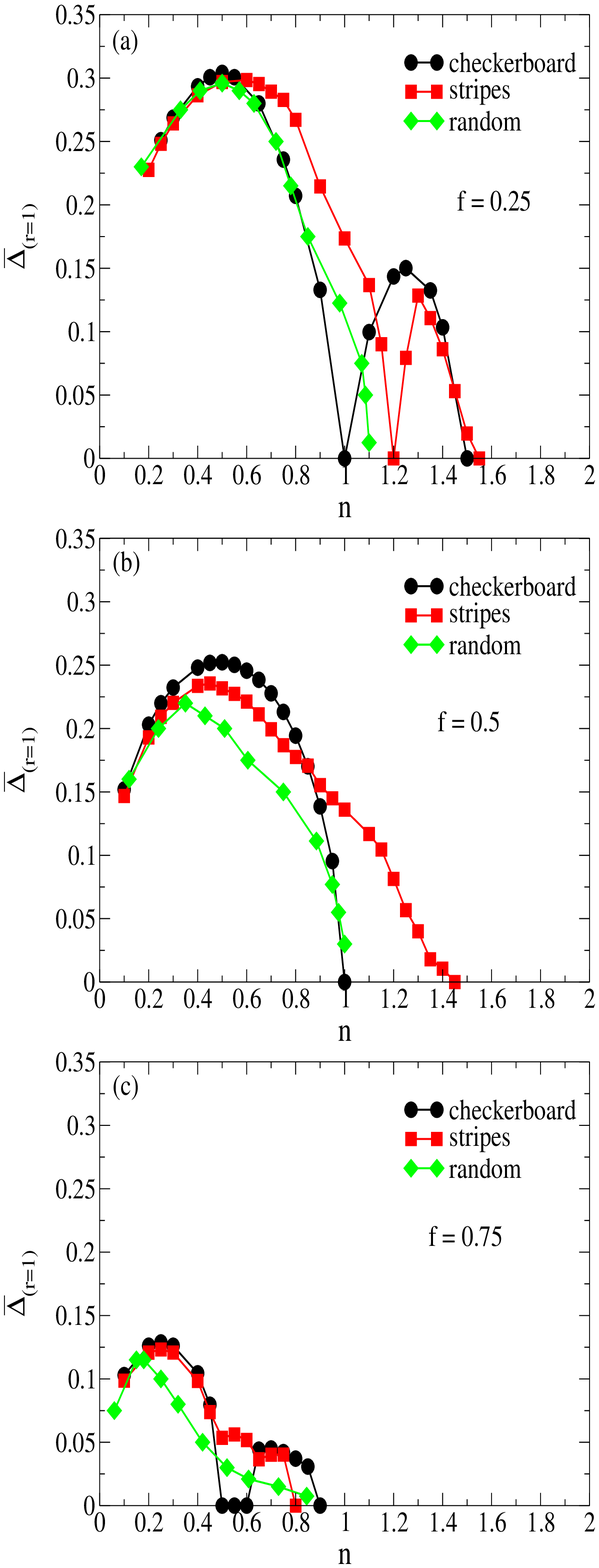}
\caption[a]{(Color online) The magnitude of the averaged order parameter
$\bar\Delta$ for the inhomogeneous patterns along the enhancement
boundary ($r=1$) in Fig. \ref{patterns} as a function of the average
electron occupation $n$ for $f=0.25$ (panel (a)), $f=0.5$ (panel (b))
and  $f=0.75$ (panel (c)).} \label{delta-r=1} \end{figure}

\par As mentioned earlier in this section, for $n<n^*$,
${\bar\Delta}_{\rm inhom}$ increases as a function  of $|\bar U|$ and
saturates for large enough $|\bar U|$ values while ${\Delta}_{\rm
uniform}$ is a consistently increasing function of $|\bar U|$ and is
symmetric with respect to $n=1$. Therefore, for $n<n^*$ as illustrated
in Fig. \ref{patterns} for $r=\bar\Delta_{\rm inhom}/\Delta_{\rm
uniform}$ at a given $n$
\begin{equation} \label{eq:r1.r2} {\bar U}_{2}> {\bar U}_{1} \Rightarrow
r({\bar U}_{2})< r({\bar U}_{1})\,.  \end{equation}  
Now since
\begin{eqnarray} \label{eq:delU1.delU2} {\bar\Delta}_{\rm inhom}({\bar
U}_{2})>{\bar\Delta}_{\rm inhom}({\bar U}_{1}) \Rightarrow \nonumber \\
{\bar\Delta}_{\rm inhom}(r({\bar U}_{2}))>{\bar\Delta}_{\rm
inhom}(r({\bar U}_{1}))\nonumber \\ {\rm for}\hspace{1.0cm}(n<n^*)\,,
\end{eqnarray}
and as a result
\begin{eqnarray} \label{eq:r=1.r>1} {\bar\Delta}_{\rm
inhom}(r=1)>{\bar\Delta}_{\rm inhom}(r>1))\nonumber \\ {\rm
for}\hspace{1.0cm}(n<n^*)\,.  \end{eqnarray}
Therefore, knowing that $r=1$ yields the largest magnitude of
${\bar\Delta}_{\rm inhom}$ that is still enhanced compared to
${\Delta}_{\rm uniform}$ when $n<n^*$, the optimum effect due to
inhomogeneity corresponds to a value of $n$ along the $r=1$ contour in
Fig. \ref{patterns}, for which ${\bar\Delta}_{\rm inhom}$ is maximized.
Fig. \ref{delta-r=1} demonstrates the variation of ${\bar\Delta}_{\rm
inhom}(r=1)$ as a function of $n$ for different $f$ values. In panel (a)
corresponding to $f=0.25$, all three different inhomogeneous patterns
yield the maximum ${\bar\Delta}_{\rm inhom}(r=1)$ within the range of
$n=0.5$ to $0.75$. In panel (b) for $f=0.5$, the maxima are closer to
$n=0.5$ while in panel (c) for $f=0.75$ they are around $n=0.25$. These
results indicate that apparently the optimum value for the doping of
electrons in these inhomogeneous systems is  $n_{opt}\sim 1-f=n^*/2$.
In strong coupling this density corresponds to leaving the free sites
empty and singly occupying the attractive ones.  By comparing this
behavior  with the uniform system for which, due to particle-hole
symmetry, $n_{opt}=1$ we can understand why $n_{opt} \sim n^*/2$.

 Also, one observes in Fig. \ref{delta-r=1} that by further diluting the
interacting sites in a lattice, the magnitude of ${\bar\Delta}_{\rm
inhom}$ at $n_{opt}$ decreases. %

\par We conclude in this section that by further diluting the density of
interacting sites in a lattice while maintaining the average pairing
energy per site constant at $T=0$, the average order parameter
${\bar\Delta}_{\rm inhom}$ may enhance. This enhancement results from
the proximity effect in the non-interacting sites due to their
interacting neighbors leading to a larger average order parameter
compared to the uniform lattice and in many respects is independent of
the particular inhomogeneous pattern. Superconductivity in an
inhomogeneous lattice of interacting sites gradually vanishes above
$n=n^*$ and for large enough $|\bar U|$ values at $n=n^*$ we have an
insulating phase as a result of a charge ordered phase formation. For
larger $f$ values, the enhancement ratio $r=\bar\Delta_{\rm
inhom}/\Delta_{\rm uniform}$ will be larger for small values of $|\bar
U|$ and $n$. However, the enhancement of $\bar\Delta_{\rm inhom}$ occurs
at smaller values of $n$. We also find an optimum value of $n_{opt}\sim
1-f=n^*/2$ for the largest enhanced $\bar\Delta_{\rm inhom}$ for a
system in the presence of inhomogeneity.   
\begin{figure} \includegraphics[width=2.6in]{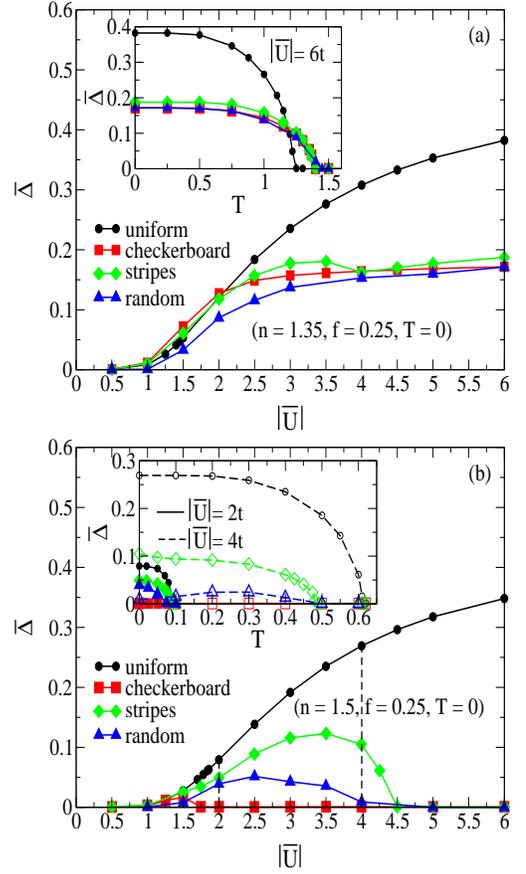}
\caption[a]{(Color online) Panel (a): (main) The variation of the
averaged order parameter $\bar\Delta$ for the uniform and three
different inhomogeneous patterns of checkerboard, stripes and random at
$f=0.25$, $n=1.35 < n^*$ and $T=0$ as a function of $|\bar U|$. The
inset presents the evolution of these order parameters as a function of
temperature for ${|\bar U|}=6t$ (The largest at $T=0$). Panel (b):
(main) The same results as in panel (a) for $n=1.5=n^*$ and $T=0$. The
inset shows the evolution of order parameters against temperature for
two different values of ${|\bar U|}=2t$ (solid line with filled symbols)
and $4t$ (dashed line with open symbols) as indicated in the $T=0$
results by the dashed lines.} \label{delta.vs.U.and.T.f0.25}
\end{figure}
\begin{figure} \includegraphics[width=2.6in]{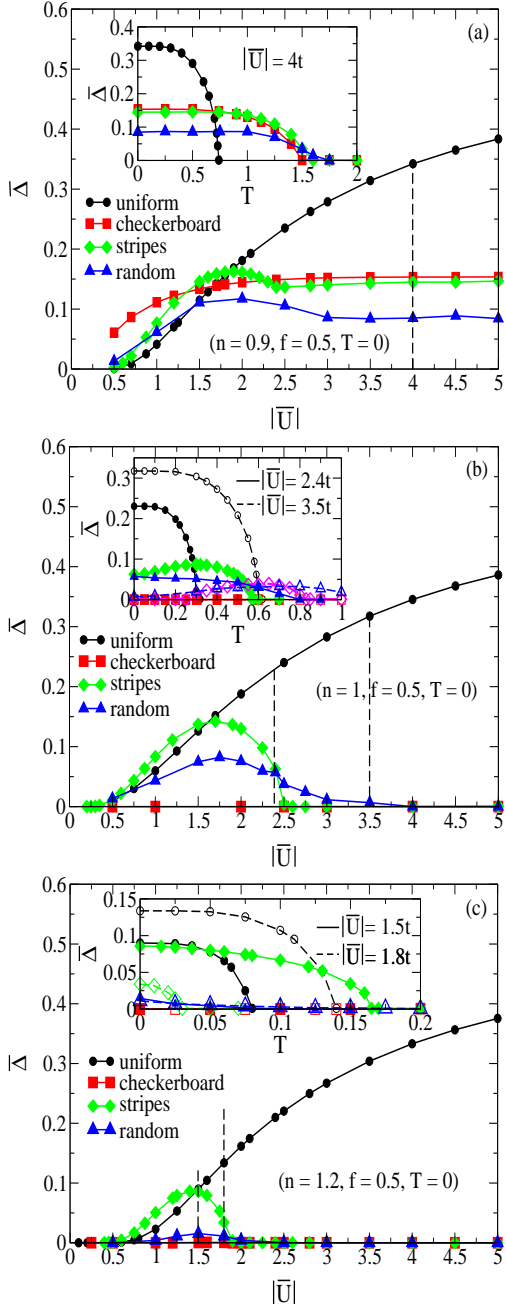}
\caption[a]{(Color online) Panels (a), (b) and (c) refer to $f=0.5$ and
$n=0.9<n^*$, $n=1 = n^*$ and $n=1.2 > n^*$ respectively. Two different
values of $|\bar U|$ have been chosen in panels (b) and (c) at finite
temperature as shown by the dashed lines in the $T=0$ results. In panel
(b), for better visibility, results for stripes at $|\bar U|=3.5t$
(dashed line with open diamonds) are shown in magenta inside the inset.}
\label{delta.vs.U.and.T.f0.5} \end{figure}
\begin{figure} \includegraphics[width=2.6in]{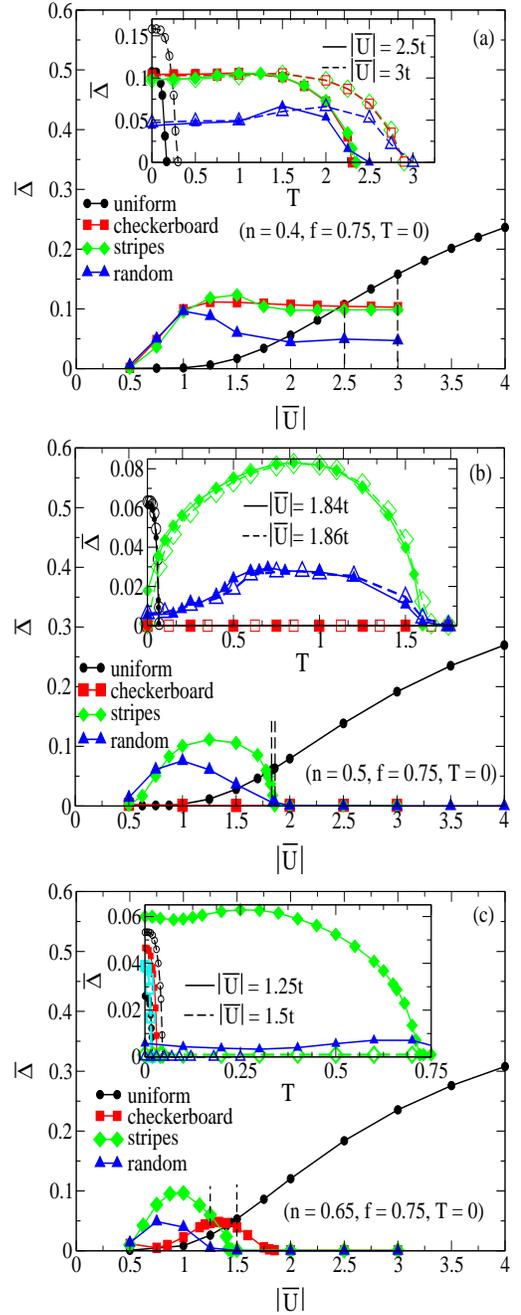}
\caption[a]{(Color online) Panels (a), (b) and (c) refer to $f=0.75$ and
$n=0.4 < n^*$, $n=0.5 = n^*$ and $n=0.65 > n^*$ respectively. Two
different values of $|\bar U|$ have been chosen in all panels as shown
by the dashed lines in the $T=0$ results. In panel (c), results for
checkerboard at $|\bar U|=1.5t$ (dashed line with open diamonds) are
shown in cyan inside the inset for better visibility.}
\label{delta.vs.U.and.T.f0.75} \end{figure}
\begin{figure}
\includegraphics[width=2.4in]{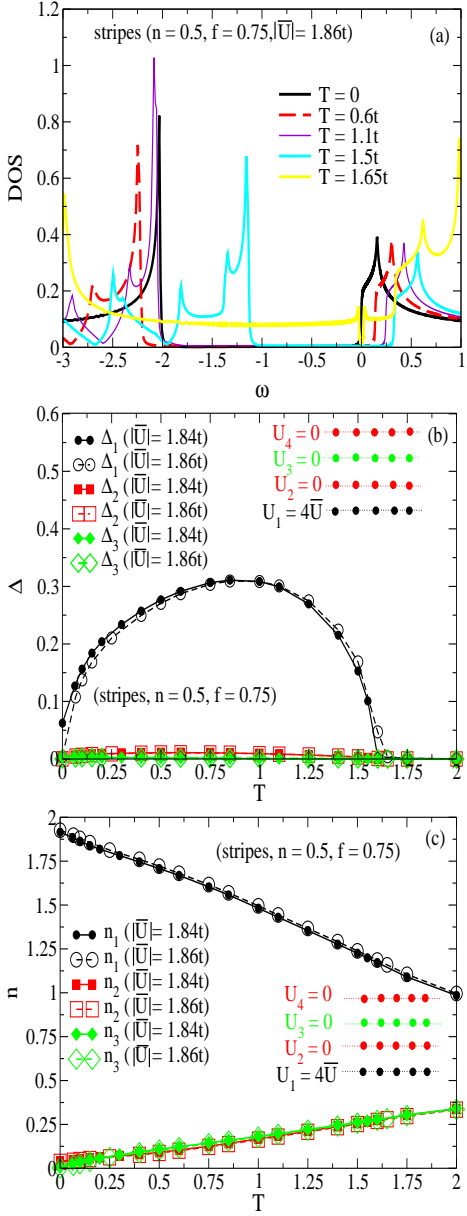}
\caption[a]{(Color online)  Anomalous behavior of the averaged order
parameter $\bar\Delta$ as a function of temperature $T$ at $f=0.75$ and
$n=0.5$ as presented in Fig.\ref{delta.vs.U.and.T.f0.75}, panel (b)
(inset), for stripes.  Panel (a) illustrates the evolution of the DOS as
a function of temperature for $|\bar U|=1.86t$. Panel (b)  demonstrates
how the local $\Delta_i$ on any of the individual four sites inside the
block shown in  Fig.\ref{patterns}, panel (b) vary as a function of
temperature $T$. Panel (c): The evolution of the local occupation number
$n_i$ on any of the individual four sites inside the block shown in
Fig.\ref{patterns}, panel (b) as a function of temperature $T$.}
\label{anomaly.stripes.n0.5.f0.75} \end{figure}
\begin{figure}
\includegraphics[width=2.4in]{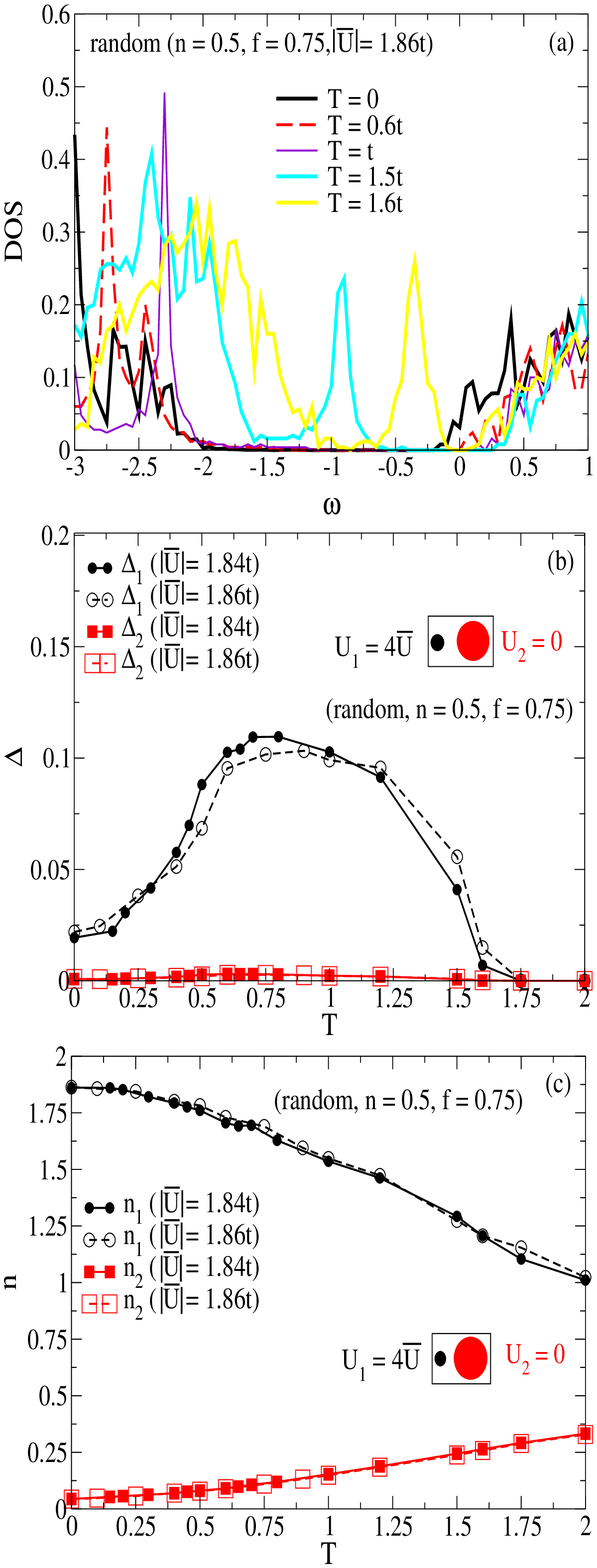}
\caption[a]{(Color online) Same as Fig. 10, but for the random pattern.}
\label{anomaly.random.n0.5.f0.75} \end{figure}
\section{Results at Finite $T$} \label{sec:results}

\par The SC transition temperature $T_{c}$ of a lattice with an
inhomogeneous pattern of interacting sites can also be larger compared
to a uniform interaction distribution on the same lattice. In this
section, we investigate the conditions under which inhomogeneity in any
form can lead to the enhancement of $T_{c}$ as a function of $|\bar U|$
and $n$ at different concentration values $f$. In Fig.
\ref{delta.vs.U.and.T.f0.25}, panel (a) presents the variation of the
average order parameter $\bar\Delta_{\rm inhom}$ and $\Delta_{\rm
uniform}$ as functions of $|\bar U|$ for $f=0.25$, $n=1.35<1.5=n^*$, and
$T=0$. We pick the value of $|\bar U|=6t$, the largest in panel (a), at
$T=0$ and plot both $\bar\Delta_{\rm inhom}(T)$ and $\Delta_{\rm
uniform}(T)$ as functions of $T$ in the inset inside panel (a). As seen
in panel (a), for $|\bar U|=6t$, $\Delta_{\rm uniform}(T=0)$ has already
exceeded all its inhomogeneous counterparts appreciably. Nevertheless,
in the inset inside the same panel, $T_{c}$ for the inhomogeneous
patterns are still larger than their uniform pattern counterpart at
$|\bar U|=6t$ indicating the strong enhancement of $T_{c}$. 

For the uniform pattern at all values of $|\bar U|$ and $n$, we find
$T_{c}$ in good agreement with the BCS prediction, $k_{b}T_{c}\approx
(\Delta(0)|\bar U|)/1.76$, as expected from our mean field approach. In
panel (b) in Fig.  \ref{delta.vs.U.and.T.f0.25} for $n=1.5=n^*$,
however, for $|\bar U|=2t$  for which $\Delta_{\rm uniform}(T=0)$ is
slightly larger than $\bar\Delta_{\rm inhom}(T=0)$, we find that $T_{c}$
for all inhomogeneous patterns (except the checkerboard whose
$\Delta_{\rm inhom}(T=0)=0$ at $|\bar U|=2t$) are also slightly larger
than the uniform pattern $T_{c}$ as shown in the inset of the same
panel. At $|\bar U|=4t$ for which $\Delta_{\rm uniform}(T=0)$ becomes
noticeably larger than $\bar\Delta_{\rm inhom}(T=0)$, as presented in
the inset inside panel (b), $T_{c}$ for the uniform pattern also becomes
larger than its inhomogeneous pattern counterparts. In other words, the
enhancement of $T_{c}$ is rather weak when $n\geq n^*$ compared to $n <
n^*$ and ceases to persist as $|\bar U|$ increases. In panel (b), the
checkerboard pattern has a vanishing average order parameter at both
$T=0$ and finite $T$. For stripes, $\bar\Delta_{\rm inhom}$ starts with
a finite value at $T=0$ and gradually vanishes as $T$ increases. For the
random pattern, $\bar\Delta_{\rm inhom}$ starts at a value very close to
zero at $T=0$. However, at $|\bar U|=4t$ as $T$ increases, there is a
slight rise in the magnitude of $\bar\Delta_{\rm inhom}$ over some
finite temperature window before it completely vanishes at high enough
$T$.     

\par In Fig. \ref{delta.vs.U.and.T.f0.5}, panel (a) presents similar
results for $f=0.5$ at $n=0.9<n^*$ and $|\bar U|=4t$. As illustrated in
the inset of the same figure, even at $|\bar U|=4t$ for which
$\bar\Delta_{\rm inhom}(T=0)<\Delta_{\rm uniform}(T=0)$, all
inhomogeneous patterns lead to larger $T_{c}$ compared to the uniform
pattern. In panel (b), for $n=1=n^*$, at $|\bar U|=2.4t$,
$\bar\Delta_{\rm inhom}(T=0)<\Delta_{\rm uniform}(T=0)$.  However, except
for the checkerboard, we still find an increase in $T_{c}$ due to
inhomogeneity. Similar to Fig. \ref{delta.vs.U.and.T.f0.25} panel (b),
for both striped and random patterns, we also find a gradual increase in
$\bar\Delta_{\rm inhom}(T)$ as $T$ increases before it totally vanishes
at high enough $T$. The enhancement of $T_{c}$ continues to persist up
to $|\bar U|=3.5t$, although  $\bar\Delta_{\rm inhom}(T=0)\approx 0$ for
all inhomogeneous patterns, due to the gradual increase of
$\bar\Delta_{\rm inhom}(T)$ with temperature.  Panel (c) corresponds to
$n=1.2>n^*$ and it can be observed that for $|\bar U|=1.5t$, for stripes
only, $T_{c}$ is increased. However, at slightly larger $|\bar U|=1.8t$,
$T_{c}$ for the uniform pattern significantly dominates  the
inhomogeneous pattern ones and no gradual increase in $\bar\Delta_{\rm
inhom}(T)$ occurs as $T$ increases.

\par Fig. \ref{delta.vs.U.and.T.f0.75}(a) has $f=0.75$
and $n=0.4<n^*$.  Similar to the behavior observed
for $f=0.25$ and $0.5$ when $n<n^*$, up to the largest value of $|\bar
U|=3t$, inhomogeneity significantly
increases $T_{c}$ as illustrated in the inset of the same panel. 
This occurs despite
the fact that $\bar\Delta_{\rm inhom}(T=0)<\Delta_{\rm uniform}(T=0)$.
For   $n=0.5=n^*$, even at $|\bar U|=1.86t$ where $\bar\Delta_{\rm
inhom}(T=0)\approx 0$, for both the striped and random patterns we again
find an increase in $T_{c}$ due to inhomogeneity.  In this case, the
gradual increase in $\bar\Delta_{\rm inhom}(T)$ as a function of $T$ is
further pronounced until $T_{c}$ totally vanishes for these two patterns
at high enough $T$. Panel (c) corresponds to $n=0.65>n^*$ and similar to
$f=0.25$ and $f=0.5$ cases when $n>n^*$, for large enough $|\bar
U|=1.5t$ (slightly larger than $|\bar U|=1.25t$ in the inset)
inhomogeneity no longer yields larger $T_{c}$ compared to the uniform.
(See inset.) 

\par The anomalous increase of $\bar\Delta_{\rm inhom}$ as a function of
$T$ at $n=n^*$ for $f\geq 0.5$ is an actual feature and is believed to
be related to the gradual destruction of the charge ordered phase due to
temperature leading to an intermediate SC phase.  In Fig.
\ref{anomaly.stripes.n0.5.f0.75} corresponding to Fig.
\ref{delta.vs.U.and.T.f0.75}, panel (b) for $n=0.5$, $f=0.75$ and $|\bar
U|=1.86t$, panel (a) presents the DOS results at several $T$ values
within the temperature window of the inset in Fig.
\ref{delta.vs.U.and.T.f0.75} panel (b). At $T=0$, the gap in the DOS is
barely non-zero at $\omega=0$ (Fermi energy) indicating a charge ordered
phase. By increasing $T$ to $0.6t$ the gap widens towards
superconductivity consistent with the behavior shown inside the inset in
Fig.  \ref{delta.vs.U.and.T.f0.75}, panel (b). By further increasing
$T$, the gap begins to shrink due to the destruction of
superconductivity by temperature until it entirely vanishes at $T=1.65t$
in agreement with the results inside the inset in Fig.
\ref{delta.vs.U.and.T.f0.75}, panel (b). 

Panel (b) in Fig.  \ref{anomaly.stripes.n0.5.f0.75} illustrates the
variation of the local order parameter on all four different types of
sites for stripes at $f=0.75$. Only
$\Delta_{\rm 1}$ along the line of interacting sites varies appreciably
as $T$ increases.  Panel (c)  of Fig. \ref{anomaly.stripes.n0.5.f0.75}
shows the local occupation numbers on all the four different types of
sites for stripes at $f=0.75$ (Fig.  \ref{patterns}, panel (b)). By
increasing $T$, while the occupation of the interacting sites gradually
drops from $n=2$ to $n\approx 1$, an indication of the destruction of
the charge ordered phase, the occupation of the non-interacting sites
grows leading to additional mobility and overlap of the Cooper pairs and
therefore enhancing $T_{c}$. 

Fig. \ref{anomaly.random.n0.5.f0.75}, panel (a) shows the DOS results
for the random pattern for $n=0.5$, $f=0.75$ and $|\bar U|=1.86t$
depicted in  Fig.  \ref{delta.vs.U.and.T.f0.75}, panel (b). Similar to
stripes, over the window of temperature studied in panel (a), the gap in
the DOS initially grows up to $T\approx t$ and then gradually drops 
to zero slightly beyond $T=1.6t$ to indicate the formation of a metallic
phase as the temperature destroys superconductivity. Panel (b) in Fig.
\ref{anomaly.random.n0.5.f0.75} also shows significant variation of
the local order parameter only on the interacting sites on average.
Hence, similar to stripes, temperature acts against the more localized
charge ordered phase allowing for the Cooper pairs to move and overlap
more freely and consequently the enhancement of $T_{c}$. Panel (c) also
confirms that the charge ordered phase is indeed destroyed by
temperature allowing the non-interacting sites to become more occupied
resulting in an intermediate SC phase. Metallic behavior is
established at sufficiently high temperature. 

It is also noteworthy that this anomalous increase in $\bar\Delta_{\rm
inhom}$ was not observed for the checkerboard pattern for which there is
no superconductivity at any temperature when $n=n^*$. One possible
explanation is that due to the            particular  geometry for the
checkerboard, localized Cooper pairs are further apart from one another
compared to  the striped and random patterns. As mentioned earlier,
when $n=n^*$ we have two electrons per  interacting site.  For the
checkerboard lattice, this leads to a pair localized in the interacting
sites with empty  nearest neighboring sites. Therefore, the effect of
temperature is minor in causing further overlap  among the pairs before
it totally destroys them (especially at $f=0.5$ and $0.75$ as shown in
Fig. \ref{patterns}, panel(a)).   

\par A similar increase in $T_c$ upon introducing a checkerboard pattern
is found in the MCMF calculations as well, arising from the loss of
long-range phase coherence. This is particularly significant because the
MCMF incorporates the subtle nature of the SC transition in $2D$
discussed earlier. We have also independently confirmed that our
conclusions and arguments equally apply for a model with
nearest-neighbor attraction, leading to a $d$-wave SC close to
half-filling, which reflects the cuprates' phenomenology more
truthfully.\cite{olderwork,mayr}

\par We have found that upon introducing
inhomogeneity into the pattern of interacting sites on a lattice. 
$T_c$ can be increased over a wide range of
$|\bar U|$ as long as $n\leq n^*$ even if $\bar\Delta_{\rm
inhom}(T=0)<\Delta_{\rm uniform}(T=0)$. When $n>n^*$, however, for
sufficiently large $|\bar U|$, inhomogeneity acts against
superconductivity and therefore $T_{c}$ becomes smaller compared to the
uniform pattern of interacting sites. The case of $n=n^*$ is anomalous.
The charge ordered phase established at large enough $|\bar U|$ values
at $T=0$ gradually vanishes as $T$ increases.  $\Delta_{inhom}$ first
increases with increasing T, leading to a SC phase (at least for large
enough $f$ values and lower symmetrical inhomogeneity patterns), and
then vanishes, indicating a metallic state.
\section{Summary and Discussion} \label{sec:summary}

\par In summary, we have shown that for the attractive Hubbard model on
a square lattice, there is a significant range of electron doping and
interaction strength over which the average superconducting order
parameter is larger for a lattice with inhomogeneous patterns of
interacting sites than a uniform distribution of these interacting sites
at a constant interaction strength per site. We have presented the phase
diagrams for three different inhomogeneous patterns:  checkerboard,
stripes and random and also three different values for the concentration
of the non-interacting sites. Apart from a few particular features, the
overall physics illustrated in the phase diagrams is pattern
independent.  As we vary the mean interaction strength $|\bar U|$ and
the doping level $n$ at zero temperature, we have verified the existence
of at least three different phases, namely (i) superconducting, (ii)
insulating due to the charge order phases, and (iii) metallic. Our
findings and claims are strongly supported by studying the behavior of a
variety of quantities computed in this work all consistently
corroborating one another.  

\par The enhancement of the average order parameter for the
inhomogeneous interacting site patterns is due to the proximity effect,
i.e., the tunneling effect of the Cooper pairs from the interacting
sites leading to finite order parameter values on neighboring 
sites. This conclusion is supported by the effect occurring at weak
coupling, where the coherence length is large, rather than in the strong
coupling regime of preformed pairs. Agreement between the BdG results
and the MCMF calculations justifies the application and conclusions of
the BdG approach within the small $|\bar U|$ regime. Our calculations
also clearly confirm that an inhomogeneous interaction potential can
lead to the increase in the phase transition temperature $T_c$ over a
wide range of $n$, and $|\bar U|$ for various $f$ values.
Counterintuitively, as long as $n$ is less than or equal to twice the
fraction of interacting sites, this increase in $T_c$ continues even for
values of $|\bar U|$ for which the order parameter is larger for the
uniform pattern than for inhomogeneous patterns at $T=0$. 

\par One possible explanation takes into account that in this weak coupling
parameter regime, $T_c$ is a supralinearly increasing function of $U$.
In such a case, it may be that in the inhomogeneous system the sites
with larger $U$ produce a nonlinear enhancement relative to $T_{c}$ of
the uniform system and, through the proximity effect, drag the
non-interacting sites along with them.  This trend changes when $n$
exceeds twice the number of interacting sites ({\i.e.} some electrons
must occupy non-interacting sites), for which at large enough $|\bar U|$
values inhomogeneity fails to increase $T_c$ over that of the uniform
pattern. The $n=2(1-f)=n^*$ case for sufficiently large $f$ values and
less symmetric inhomogeneous patterns (such as stripes and random as
opposed to the checkerboard) is anomalous as it shows the enhancement of
$\bar\Delta_{\rm inhom}(T)$ as temperature increases. 

\par It is even more surprising to find that a system which is
non-superconducting (charge ordered insulator) at $T=0$ can become
superconducting upon increasing $T$ for a finite window of temperature
before turning metallic. This anomalous
behavior was shown to be related to a crossover from a charge ordered
insulating phase for $n=n^*$ at large enough $|\bar U|$ values to an
intermediate SC phase upon increasing $T$ before entering the metallic
phase at sufficiently large $T$.

\par We wish to emphasize that in this article we have focused on the enhancement (or not) of the pairing amplitude (our $\Delta$ defined in section \ref{sec:formalism}), rather than its product with the local interaction strength $-|U_i|$ which is more directly related to the gap but contains less information, and gives less insight, because it automatically vanishes on any site without interaction. Thermodynamic measurements would probe quantities which include the energy scale, such as the specific heat or superfluid rigidity which our results may not have direct implications to. However, for an inhomogeneous system being a mixture of different phases, defining an average SC gap is not trivial. Thus, the thermodynamic properties of inhomogeneous superconductors will not necessary exhibit the same behavior as their homogeneous counterparts. It has been shown that the rise of the specific heat in inhomogeneous superconductors obeys a power law behavior as opposed to exponential in homogeneous ones using the attractive Hubbard model with random interacting sites. \cite{litak2} Also, the superfluid density and stiffness in general decrease due to the presence of disorder. \cite{paramekanti}  Nevertheless, lower superfluid density does not necessarily lead to lower $T_c$ as according to the Anderson theorem, a non magnetic impurity should not affect the $T_c$ and therefore thermodynamic properties of a s-wave superconductor. We have presented clear evidence for the enhancement of $T_c$ which does have a direct experimental implication. Thermodynamic properties of inhomogeneous superconductors are very rich in physics and a great deal of contributions and new ideas in this area are yet to appear.

\par While the attractive Hubbard Hamiltonian obviously does not
incorporate many of the features of high $T_c$ superconductors (notably
the symmetry of the pairing), the model has been shown to provide useful
insight into some of their phenomenology, for example the
spin-gap.\cite{randeria94} It is therefore tempting to speculate that
our results concerning inhomogeneity may have similar connections.
Specifically, earlier ARPES data \cite{yoshida03} suggests that the
underdoped phase of LSCO (La$_{2-x}$Sr$_{x}$CuO$_{4}$) consists of SC
clusters, embedded in the AF host. In such a system, inhomogeneous gaps
appear naturally and our results here indicate that the SC transition is
in fact determined by the largest gap values rather than the much
smaller gaps found at phase boundaries, as one might naively think. This
renders the SC phase more stable than it would otherwise be, and also
simplifies the description of these systems.

It is worth emphasizing that in most situations, inhomogeneities reduce
values of order parameters \cite{chatterjee} and critical temperatures,
even when comparisons are made, as they are in this article, to
homogeneous systems with the same average value of all parameters. This
is true, for example, for classical site diluted Ising models, where the
ferromagnetic $J$ is increased to compensate for absent sites, and
quantum models like the boson Hubbard model where random chemical
potentials monotonically decrease and ultimately destroy
superfluidity.\cite{fisher89,scalettar91} An exception is the increase
of $T_{\rm Neel}$ by randomness reported in DMFT studies of the
repulsive model \cite{ulmke95} and recently, the SC $T_{c}$ in $XY$
model Hamiltonian with certain types of inhomogeneous patterns for the
coupling constant between spins sitting on nearest neighboring
sites. \cite{loh}.
\section{Acknowledgments} \label{sec:acknowledgments}

\par We acknowledge useful conversations with Jian-Xin Zhu and Wei Ku.
This research was supported by National Science Foundation Grant
DMR-0421810, DMR-0426826, US ONR, CNPq-Brazil and FAPERJ-Brazil. %

\end{document}